\documentclass[lettersize,journal]{IEEEtran}
\usepackage{amsmath,amsfonts}
\usepackage{algorithmic}
\usepackage{algorithm}
\usepackage{array}
\usepackage{textcomp}
\usepackage{stfloats}
\usepackage{url}
\usepackage{verbatim}
\usepackage{graphicx}
\usepackage{cite}
\usepackage{subfigure}
\usepackage{graphicx} 
\usepackage{subcaption} 
\begin{document}

\title{Enhancing Secrecy capacity with non-orthogonal artificial noise based on Pilot Information Codebook}

\author{Yebo Gu, Tao Shen*, Jian Song*, Qingbo Wang
\thanks{Y.Gu, Y.Shen, and J.Song are with the College of Information Engineering and Automation, Kunming University of Science and Technology. (email:guyebo@kust.edu.cn; shentao@kust.edu.cn; songjian@kust.edu.cn).}
\thanks{Q.Wang is with Naval Research Academy.(email:qingbo1992@163.com)}}



\maketitle

\begin{abstract}
In recent research, non-orthogonal artificial noise (NORAN) has been proposed as an alternative to orthogonal artificial noise (AN). However, NORAN introduces additional noise into the channel, which reduces the capacity of the legitimate channel (LC). At the same time, selecting a NORAN design with ideal security performance from a large number of design options is also a challenging problem. To address these two issues, a novel NORAN based on a pilot information codebook is proposed in this letter. The codebook associates different suboptimal NORANs with pilot information as the key under different channel state information (CSI). The receiver interrogates the codebook using the pilot information to obtain the NORAN that the transmitter will transmit in the next moment, in order to eliminate the NORAN when receiving information. Therefore, NORAN based on pilot information codebooks can improve the secrecy capacity (SC) of the communication system by directly using suboptimal NORAN design schemes without increasing the noise in the LC. Numerical simulations and analyses show that the introduction of NORAN with a novel design using pilot information codebooks significantly enhances the security and improves the SC of the communication system.
\end{abstract}


\section{Introduction}
With the rapid development of communication technology, there has been an increasing demand for enhanced security in data transmission, while at the same time seeking efficient and fast data transmission. Although traditional wireless communication security technology based on encryption algorithms can effectively protect wireless data transmission, it has problems such as key management, information cracking, increased transmission delay and power consumption. PLS technology protects and encrypts signals at the physical layer, which has many advantages such as low computational cost, strong network applicability, anti-eavesdropping and anti-interference. It provides a new way to achieve secure wireless data transmission.

In \cite{ref-Wyner}, Wyner presented PLS technology and introduced a PLS communication model consisting of a transmitter, a receiver and an eavesdropper. Despite an initial stagnation in the development of the fundamental theory of PLS technology, researchers began investigations by considering different channel models. In particular, the methods for calculating the SC under channel models such as the broadcast channel \cite{ref-Csiszar} and the Gaussian channel \cite{ref-gaussian} are proposed, which effectively simplify the computation of PLS properties. The limited channel degrees of freedom in early models posed a challenge to the advancement of PLS theory, resulting in a long period of limited progress. However, the advent of multiple-input multiple-output (MIMO) communication systems provided a new perspective for PLS research. Researchers subsequently derived SC calculation techniques for MIMO communication systems \cite{ref-MIMO}. These developments have created new opportunities to explore the theoretical foundations of PLS technology and related concepts.

The traditional limitations of PLS technology, which required the eavesdropping channel (EC) to be a weaker version than the LC, have been overcome with the introduction of AN technology. AN has greatly expanded the application range of PLS technology. AN can effectively weaken the EC without affecting the LC\cite{ref-AN}. The emergence of AN has attracted considerable attention in the academic community and has found applications in various communication environments. The paper \cite{ref-UAV} investigates the maximisation of the average secrecy rate in a wireless communication system supported by unmanned aerial vehicles (UAVs). The study centers around the implementation of a secure transmission protocol utilizing AN injection and the simultaneous optimization of UAV trajectory, network transmission power, and AN power allocation to enhance secrecy performance.In the paper \cite{ref-imperfect}, the impact of AN on the security of a fingerprint embedding authentication framework in the presence of imperfect CSI is investigated. The results show that AN can significantly improve security, but its benefits diminish as the quality of CSI degrades, potentially compromising key security. The paper \cite{ref-massive} investigates the performance of secrecy failures in a large-scale downlink system supported by full-duplex non-orthogonal multiple access (FD-NOMA) transmission and AN. A secure cooperative communication scheme is proposed and simulations demonstrate its superiority over alternative approaches. The influence of secrecy diversity order and power allocation optimisation on system performance is also investigated.In the paper \cite{ref-IRS}, a secure MIMO wireless communication system using AN and intelligent reflecting surfaces (IRS) is studied. The joint optimization of transmit precoding, AN covariance matrix and IRS phase shifts is pursued to maximize the secrecy rate. An efficient algorithm is proposed to solve the optimisation problem with closed-form solutions for all variables. However, previous researchers have overlooked the inherent limitations associated with AN. Under natural communication conditions, the effectiveness of AN is limited by the channel degrees of freedom. In scenarios where the channel degrees of freedom are low, the design options for AN are limited. Furthermore, when the channel degrees of freedom are zero, it becomes impossible to design AN.

To overcome the limitations of AN, NORAN has been proposed \cite{mine}. NORAN focuses on designing AN in such a way that both the LC and the EC are taken into account. This approach significantly improves the confidentiality of wireless communication systems. In contrast to conventional methods, NORAN operates in the value space of the receiving CSI, which allows for greater flexibility in its design, unrestricted by the degrees of freedom of the channel. By jointly optimising the AN for both the LC and EC, NORAN achieves higher gains in terms of SC, thereby improving the overall security of wireless communication systems.

NORAN, like other types of noise, can indeed reduce the capacity of a LC. In the context of this letter, the CSI serves as a crucial element for exploring suboptimal NORAN design schemes through the codebook. By using this CSI, researchers can efficiently search for NORAN schemes that may be sub-optimal but still offer significant benefits. Once the NORAN schemes are identified, NORAN is eliminated at the receiver to mitigate its impact on wireless communication systems. This approach allows the capacity degradation caused by NORAN to be reduced while maintaining the desired level of security in the communication system.
\section{SYSTEM MODEL AND PROBLEM FORMULATION}

\subsection*{A.System Model}

\begin{figure}[h]
	\center
	\includegraphics[width=9cm,height=6.75cm]{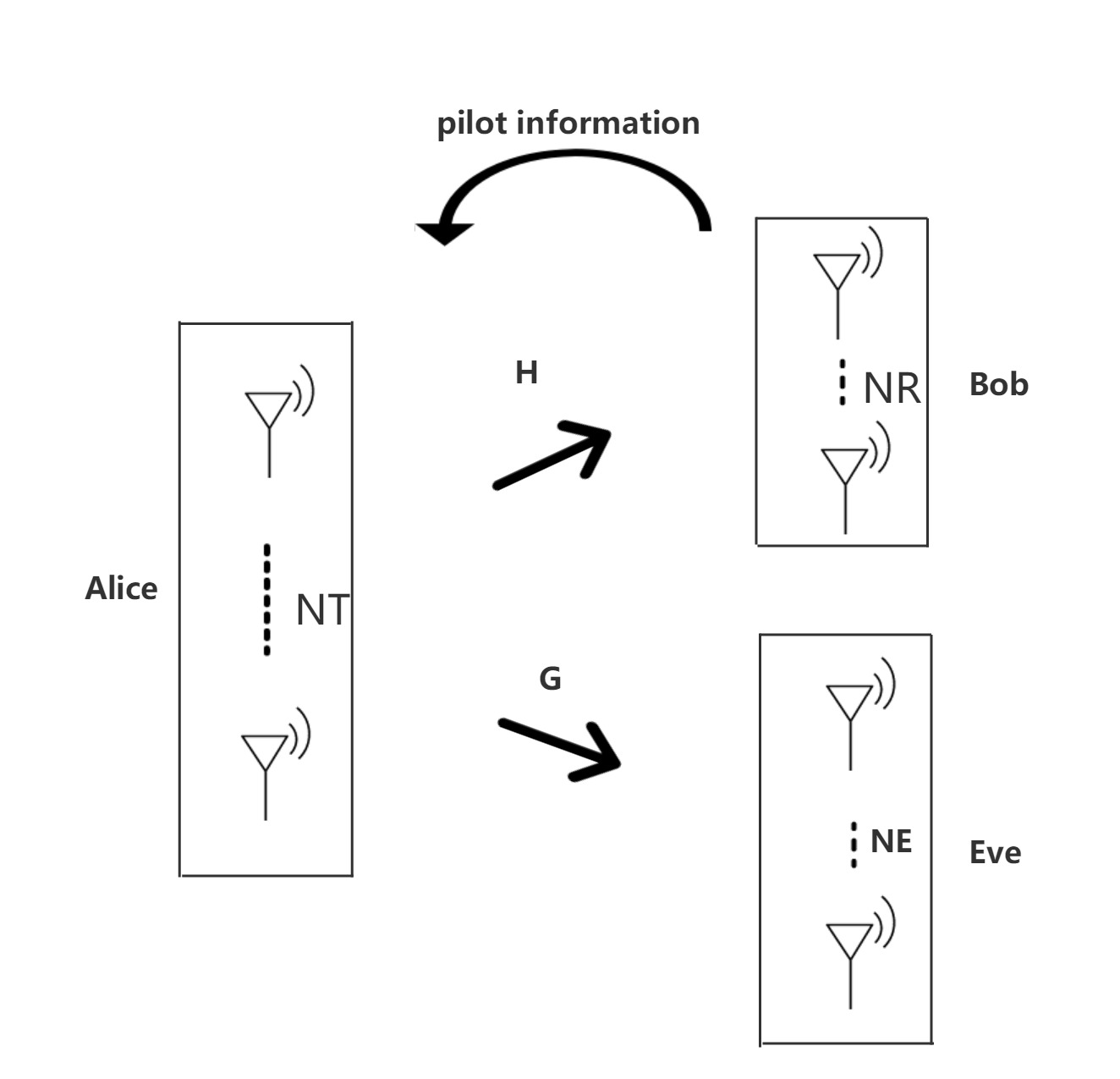}
	\caption{Wireless communication model with eavesdropper.}
\end{figure}

The communication network model shown in Fig. 1 includes a sender - Alice, a receiver - Bob, and an eavesdropper - Eve. 
Alice and Bob are connected by a bidirectional communication channel, where Bob is able to feed back CSI obtained from pilot signals to Alice. Alice has $N_T$ antennas, Bob has $N_R$ antennas and Eve has $N_E$ antennas. The LC is denoted $\mathbf H_k$ and the information sent by Alice is $\mathbf x_k$. ${\bf{x}}_k = {\bf{p}}_k{\bf{s}}_k$, where ${\bf u}_k$ is the information-bearing signal and ${\bf p}_k$ follows an independent Gaussian distribution. The variance of ${\bf u}_k$ is ${\sigma _u^2}$. ${\bf p}_k$ is chosen so that ${\mathbf H}_k{\mathbf p}_k\ne1$, $\left\| {{{\mathbf p}_k} }\right\|{\rm{ = }}1$. The NORAN is represented by $\mathbf w_k$, where ${\bf{w}}_k = {\bf{p}}_k{\bf{t}}_k$, and $\mathbf t_k$ has a variance of $\sigma _k^2$. The EC is denoted $\mathbf G_k$. The subchannels in $\mathbf H_k$ and $\mathbf G_k$ are independent and identically distributed (i.i.d.) rayleigh fading channels.  ${\mathbf H_k} \in {{\mathbb{C}^{{N_T} \times {N_R}}}}$ and ${\mathbf G_k} \in {{\mathbb{C}^{{N_T} \times {N_E}}}}$. $\mathbf n_k$ and $\mathbf e_k$ are Gaussian white noise in $\mathbf H_k$ and $\mathbf G_k$ respectively, where ${\mathbf n_{{k}}} \in {{\mathbb{C}^{N_R}}}$, ${\mathbf e_{{k}}} \in {{\mathbb{C}^{N_E}}}$, and $n \sim \mathcal{N}(0,1)$, $e \sim \mathcal{N}(0,1)$. 

After introducing NORAN into the wireless communication system, Bob and Eve receive signals as follows:
\begin{equation}{{\bf{z}}_k} = {{\bf{H}}_k}{{\bf{x}}_k} + {{\bf{H}}_k}{{\bf{w}}_k} + {{\bf{n}}_k},\label{1}\end{equation} 
\begin{equation}{{\bf{y}}_k} = {{\bf{G}}_k}{{\bf{x}}_k} + {{\bf{G}}_k}{{\bf{w}}_k} + {{\bf{e}}_k},\label{2}\end{equation}
where ${{\bf{z}}_k}$ represents the signal received by Bob , ${{\bf{w}}_k}$ represents NORAN, ${{\bf{y}}_k}$ represents the signal received by Eve.

The SC after the introduction of NORAN in the wireless communication system is
\begin{equation}
	\begin{array}{*{20}{l}}
		{{C_{wk}} = {\rm{I}}(Z;S) - {\rm{I}}(Y;S)}\\
		{\;\;\;\;\;\; \; \; = {{\log }_2}(1 + \frac{{{{\left| {{{\bf{H}}_k}{{\bf{p}}_k}} \right|}^2}\sigma _u^2}}{{{{\left| {{{\bf{H}}_k}{{\bf{p}}_k}} \right|}^2}\sigma _k^2 + \sigma _n^2}}) - {{\log }_2}(1 + \frac{{{{\left| {{\rm{ }}{{\bf{G}}_k}{{\bf{p}}_k}} \right|}^2}\sigma _u^2}}{{{{\left| {{{\bf{G}}_k}{{\bf{p}}_k}} \right|}^2}\sigma _k^2 + \sigma _e^2}}),}\label{3}
	\end{array}
\end{equation}

Since NORAN does not orthogonalise with the channel $\mathbf{H_k}$, it introduces additional noise into $\mathbf{H_k}$, reducing its capacity. At the same time, the capacity of the EC is also reduced.

\subsection*{B.Problem Formulation}

The introduction of AN technology has significantly extended the scope of PLS technology. Traditional PLS techniques are limited by the requirement that the EC must be a degraded version of the LC. However, the advent of AN allows PLS technology to overcome this limitation. AN can attenuate the EC without affecting the LC. Subsequently, researchers have explored the application of AN technology in various communication scenarios, including different channel models, communication standards and availability of CSI. However, previous studies have overlooked the inherent drawbacks of AN. In practical communication environments, the practicality of AN is quite limited. Let $\mathbf{w}_n$ represent AN, and based on the definition of AN, $\mathbf{H_k}\mathbf{w}_n=0$. The vector $\mathbf{w}_n$ corresponds to the zero solution of the homogeneous linear equation $\mathbf{H_k}\mathbf{X}=0$, where $\mathbf{X}$ is non-zero. The matrix $\mathbf{H_k}$ has the dimensions $N_R \times N_T$, where $N_R$ is the number of receive antennas and $N_T$ is the number of transmit antennas. If $\mathbf{X}$ is not zero, the rank of $\mathbf{H_k}$, denoted by rank($\mathbf{H_k}$), must satisfy rank($\mathbf{H}$) $<$ min($N_R, N_T$). This condition implies that $N_T$ and $N_R$ must be different, and min($N_R, N_T$) corresponds to the smaller value between $N_R$ and $N_T$. To implement AN in a MIMO communication system, either the $N_T$ or the $N_R$ must be reduced. However, reducing the number of antennas also reduces the capacity of the LC.

The shortcomings of the AN technology can be summarised as follows:

1) The performance of AN is limited by the requirement that it must be in the null space of $\mathbf{H_k}$, which is a special condition of the value space of the channel, and in many practical communication environments the degrees of freedom of the null space of $\mathbf{H_k}$ are small or even non-existent.

2) When incorporating AN into a wireless communication system, it is imperative to ensure that $N_R$ is strictly less than $N_T$. This condition, which is intended to facilitate the use of AN, unintentionally reduces the capacity of regular communication channels, which is contrary to the intended purpose of using AN.

3) AN can only be used in MIMO wireless communication systems, while it cannot be used in SISO wireless communication systems.

In addition, NORAN operates within the value space of the channel, making it less susceptible to variations in the channel's degrees of freedom. NORAN incorporates adaptable design solutions that can be used effectively over a wide range of channel conditions, and thus has superior universality and applicability compared to AN.

To mitigate the impact of NORAN on Bob, this letter proposes an encryption scheme based on the CSI of $\mathbf{H_k}$. In the considered model of a wireless communication system with eavesdroppers, a bidirectional link connects Alice and Bob, allowing the transmission of channel estimation information. Taking advantage of this bidirectional link, this letter presents a novel approach where the channel estimation information is used as a passphrase to generate a codebook. This codebook consists of suboptimal NORAN design schemes tailored to different channel conditions. Both Alice and Bob share the codebook.

During communication, Bob accesses the codebook while providing feedback on the CSI obtained through channel estimation. As a result, Bob can gain knowledge of the upcoming NORAN transmission from Alice and effectively eliminate its effects when receiving subsequent communication signals. The codebook is constructed by optimising the suboptimal NORAN design scheme using a continuous convex approximation algorithm with the objective function defined in Eq. (\ref{3}).

\section{PROPOSED ALGORITHM}
Because Alice and Bob use a codebook, Bob is able to cancel out NORAN when he receives the signal. Therefore, the optimisation function for NORAN is the same as for AN.

When Alice and Bob used a shared codebook, Bob effectively mitigated the presence of NORAN in the received signal by referencing the codebook. Consequently, after the codebook implementation, Bob and Eve respectively received the following signals:

\begin{equation}
{{\bf{z}}_k} = {{\bf{H}}_k}{{\bf{x}}_k} + {{\bf{n}}_k},
\end{equation}
\begin{equation}
	{{\bf{y}}_k} = {{\bf{G}}_k}{{\bf{x}}_k} + {{\bf{G}}_k}{{\bf{w}}_k} + {{\bf{e}}_k},
\end{equation}

The secrecy of the communication system after using a codebook is determined as follows:
\begin{equation}
	\begin{array}{*{20}{l}}
		{{C_{wk}} = {\rm{I}}(Z;S) - {\rm{I}}(Y;S)}\\
		{\;\;\;\;\;\; = {{\log }_2}(1 + \frac{{{{\left| {{{\bf{H}}_k}{{\bf{p}}_k}} \right|}^2}\sigma _u^2}}{{\sigma _n^2}}) - {{\log }_2}(1 + \frac{{{{\left| {{{\bf{G}}_k}{{\bf{p}}_k}} \right|}^2}\sigma _u^2}}{{{{\left| {{{\bf{G}}_k}{{\bf{p}}_k}} \right|}^2}\sigma _k^2 + \sigma _e^2}}),}
	\end{array}
\label{5a}
\end{equation}

A comparison between \ref{3} and \ref{5a} shows that the use of a codebook for NORAN elimination leads to an increase in the capacity of the LC. However, the capacity of the EC remains unaffected. Consequently, the implementation of the codebook results in an increased secure capacity of the communication system.

The use of a codebook not only facilitates the eradication of NORAN, but also provides exceptional security capabilities when extracting NORAN based on the query codebook:
\refstepcounter{equation}
\begin{align*}
	(P1)\min \;\;&{{\log }_2}(1 + \frac{{{{\left| {{{\bf{G}}_k}{{\bf{p}}_k}} \right|}^2}\sigma _u^2}}{{{{\left| {{{\bf{G}}_k}{{\bf{p}}_k}} \right|}^2}\sigma _k^2 + \sigma _e^2}}) - {{\log }_2}(1 + \frac{{{{\left| {{{\bf{H}}_k}{{\bf{p}}_k}} \right|}^2}\sigma _u^2}}{{\sigma _n^2}})\tag{4a}\label{4a}  \\
	s.t.\;\;\;&\sigma _k^2 + \;\sigma _u^2 \le P,\tag{4b} \label{4b} \\
	&\sigma _k^2 \ge 0,\tag{4c} \label{4c} \\
	&\sigma _u^2 \ge 0.\tag{4d}\label{4d}\\
	\label{4}
\end{align*}

Undoubtedly, the optimisation problem ($P1$) has a non-convex nature. In $P1$, the comparison between Eq. \eqref{4a} and Eq. \eqref{3} shows that the use of a codebook eliminates the influence of NORAN on Bob, making the impact of both non-orthogonal and orthogonal AN on Bob negligible. Nevertheless, it should be emphasized that the available options for AN are significantly more limited compared to NORAN. Consequently, under the constraint of a fixed transmit power, the potential increase in the SC of a wireless communication system with AN is significantly limited compared to that achievable with NORAN. By exploiting the codebook, NORAN effectively avoids the drawback of amplified receiver noise and can completely replace AN. In eq. \eqref{4a}, the convexity of the function ${{{\log }_2}(1 + {{{{\left| {{{\bf{G}}_k}{{\bf{p}}_k}} \right|}^2}\sigma _u^2}}/({{{{\left| {{{\bf{G}}_k}{{\bf{p}}_k}} \right|}^2}\sigma _k^2 + \sigma _e^2}}))}$ can be readily established under the constraints \eqref{4b}, \eqref{4c}, and \eqref{4b}. Additionally, ${{{\log }_2}(1 + {{{{\left| {{{\bf{H}}_k}{{\bf{p}}_k}} \right|}^2}\sigma _u^2}}/{{\sigma _n^2}})}$ represents a concave function.

The optimisation problem $P1$ can be categorised as a concave-convex program, making it amenable to decomposition into a concave function and a convex function. The algorithmic approach is to fix the convex function, optimise the concave function, and then fix the concave function while optimising the convex function. This iterative process continues until convergence to a stationary point is achieved.

To convience the disscussion, $f_{cvx}(x)$ represents ${{{\left| {{{\bf{G}}_k}{{\bf{p}}_k}} \right|}^2}\sigma _u^2}/({{{{\left| {{{\bf{G}}_k}{{\bf{p}}_k}} \right|}^2}\sigma _k^2 + \sigma e^2}}))$, and $f{cve}(x)$ represents ${{{\log }_2}(1 + {{{{\left| {{{\bf{H}}_k}{{\bf{p}}_k}} \right|}^2}\sigma _u^2}}/{{\sigma _n^2}})}$.A sequence ${x_r}$ is generated by solving the following equation:
\begin{equation}
	\nabla {f_{cvx}}({x_{r + 1}}){\rm{ }} =  - \nabla {f_{cve}}({x_r}),
\end{equation}
which is equivalent to
\begin{equation}
	{x^{r + 1}} = \arg \;\mathop {\min }\limits_x g(x,{x^r}),
\end{equation}

The function $g(x,{x^r}) \buildrel \Delta \over = {f_{cvx}}(x) + {(x - {x^r})^T}\nabla {f_{cve}}({x^r}) + {f_{cve}}({x^r})$ is a tight convex upper-bound of the objective function $f(x)$. Under certain assumptions, the convergence of $g(x,x^r)$ can be guaranteed. 

The approximate transformation of equation (4a) is performed using formula (6). Then, the problem P1 is optimised using the CVX toolbox in Matlab to obtain a suboptimal solution, denoted as $\sigma _{ko}^2$, for the NORAN variance. Based on this suboptimal solution $\sigma _{ko}^2$, the suboptimal design scheme $\sqrt {\sigma _{ko}^2} {\bf{I}}$ for NORAN is obtained. ${\bf{I}}$ is a ${N_T} \times 1$ unit vector.
\section{NUMERICAL RESULTS}

\begin{figure*} 
	\centering
	
	\subfigure[The SC of the communication system changes after introducing NORAN.]{
		\begin{minipage}[b]{0.45\linewidth} 
			\centering
			\includegraphics[width=\textwidth]{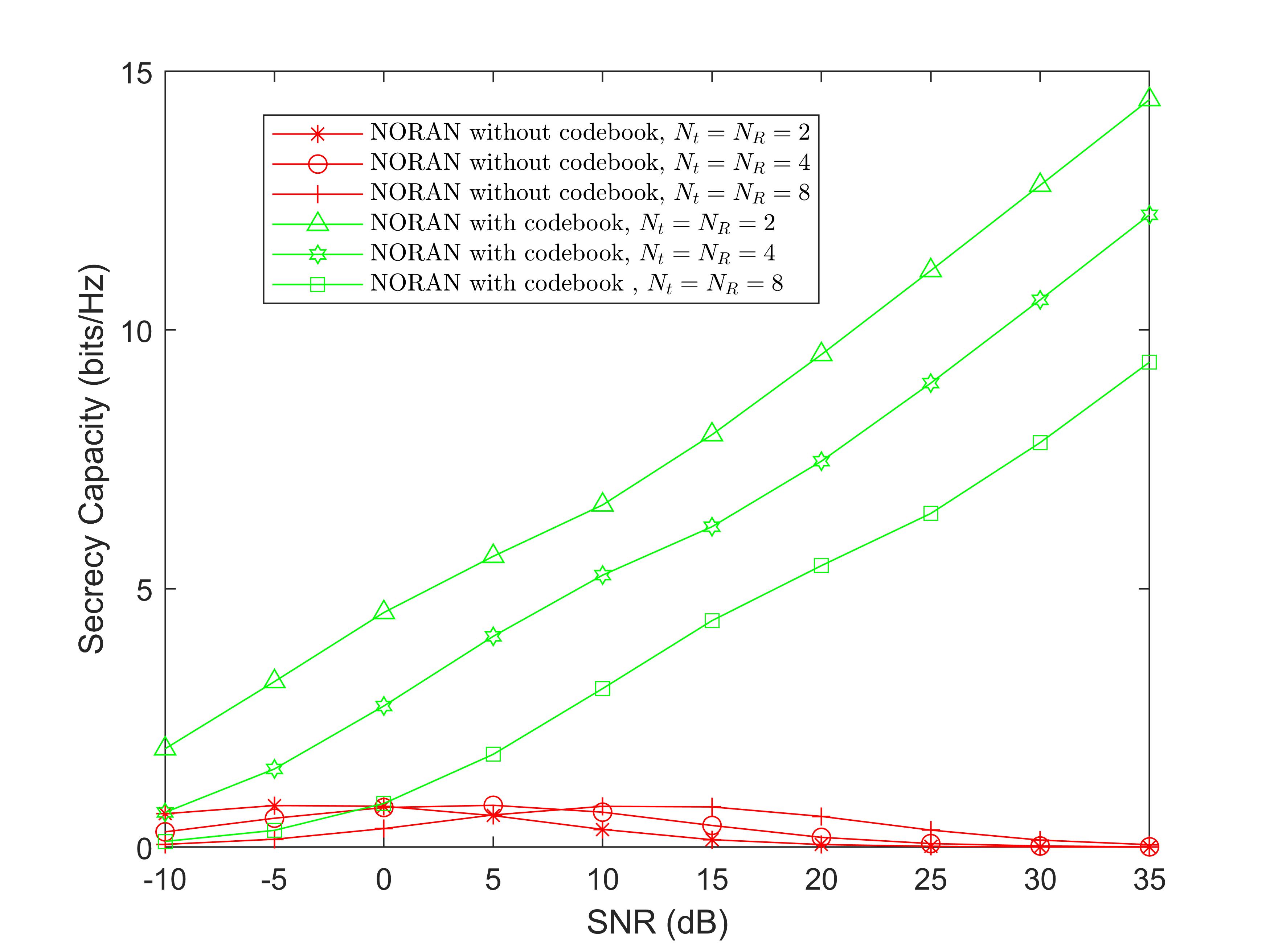} 
		\end{minipage}
	}
	\hfill 
	\subfigure[The changes in the eavesdropper's BER after introducing NORAN]{
		\begin{minipage}[b]{0.45\linewidth} 
			\centering
			\includegraphics[width=\textwidth]{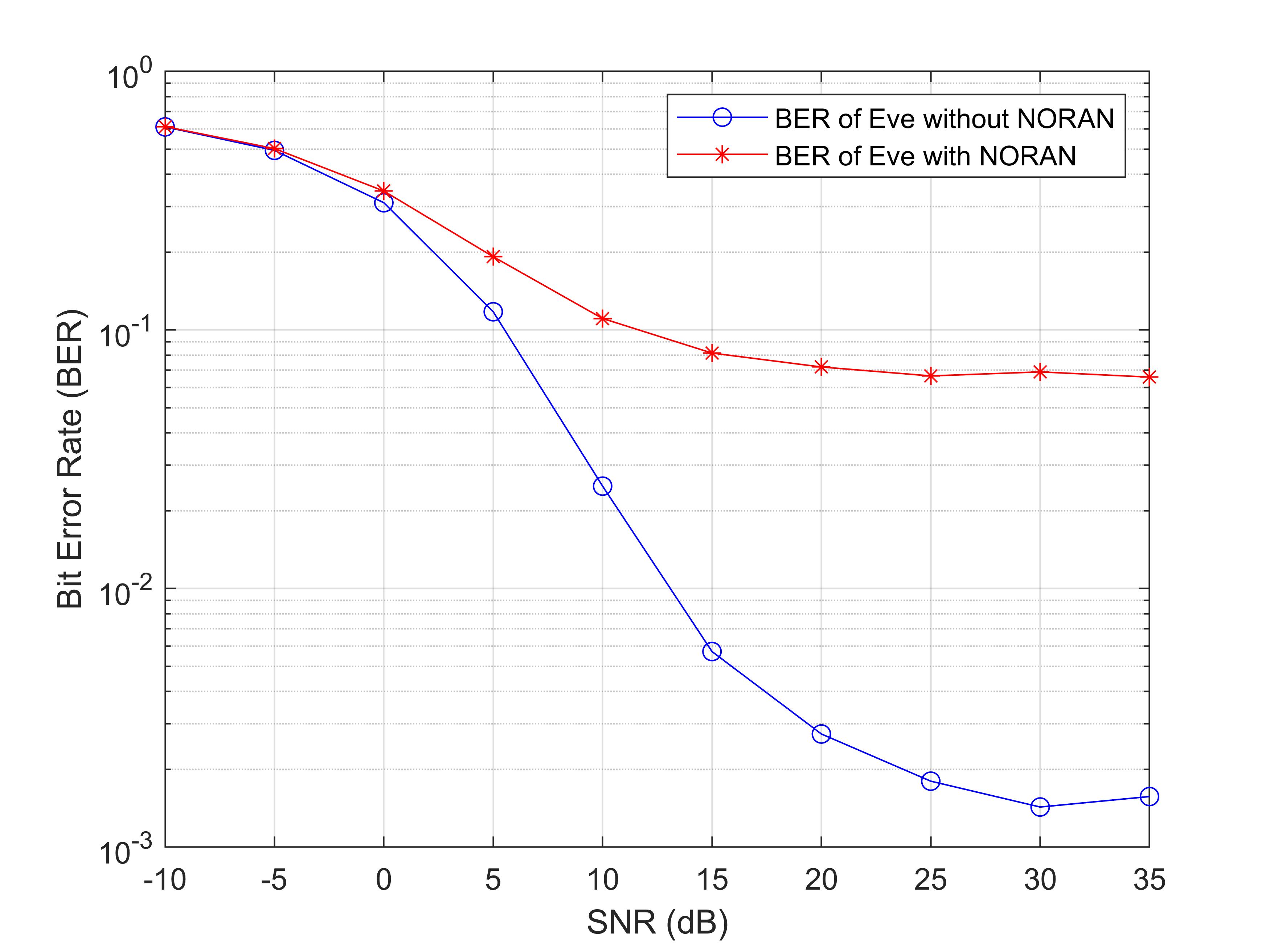} 
		\end{minipage}
	}
	%
	\caption{The changes in the security performance of the communication system after introducing NORAN.}
	\label{fig:subfigures1}
\end{figure*}
\begin{figure*} 
	\centering
	
	\subfigure[The changes in the SC of the LC with respect to SNR after introducing NORAN.]{
		\begin{minipage}[b]{0.45\linewidth} 
			\centering
			\includegraphics[width=\textwidth]{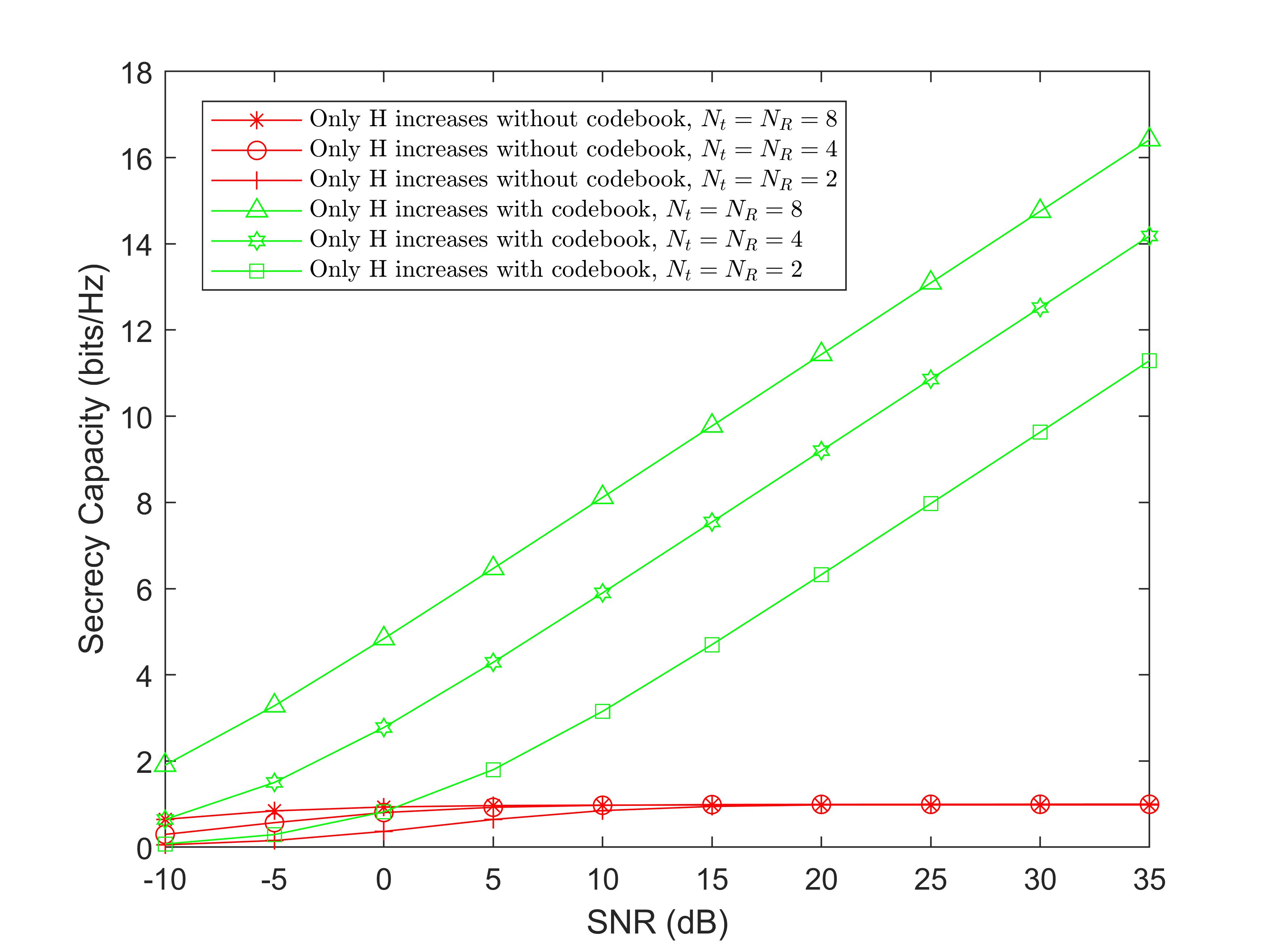} 
		\end{minipage}
	}
	\hfill 
	\subfigure[The changes in the SC of the EC with respect to SNR after introducing NORAN.]{
		\begin{minipage}[b]{0.45\linewidth} 
			\centering
			\includegraphics[width=\textwidth]{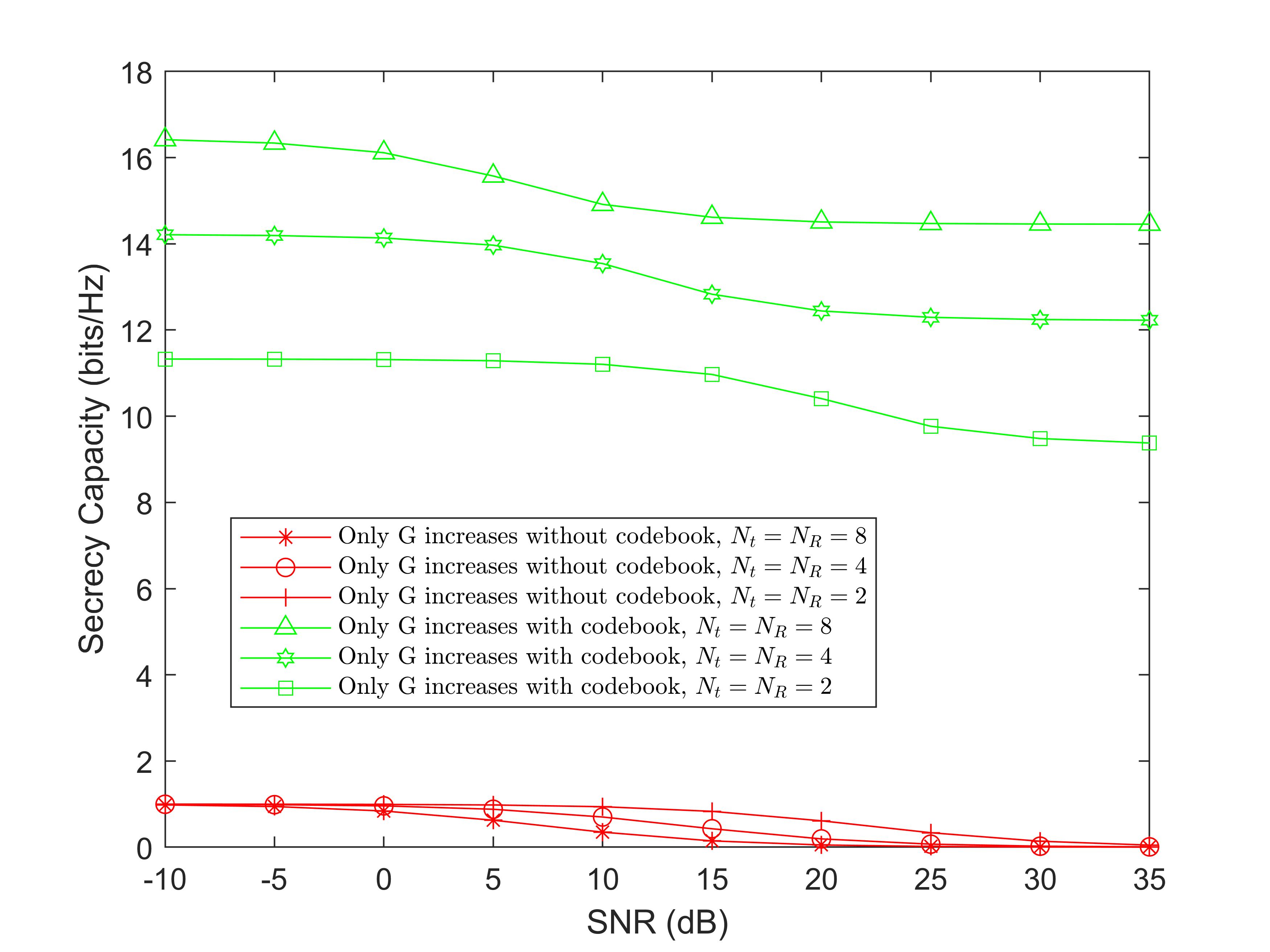} 
		\end{minipage}
	}
	
	\caption{The changes in the SC of both the LC and the EC with respect to SNR after introducing NORAN.}
	\label{fig:subfigures2}
\end{figure*}

We analyse the changes in secrecy and bit error rate (BER) of a communication system after the introduction of NORAN. By simulating the effects of various disturbances and attacks in a real communication environment, we evaluate the secure performance of the communication system with the inclusion of NORAN.

In order to assess the changes in the SC of the communication system at different signal-to-noise ratios (SNR), we measured the data rates at different SNR in our experiments. At each level of SNR, we recorded the maximum data rate that the system could reliably transmit.

Based on the experimental results, as shown in Fig. 2(a), we observed that the SC of the communication system increased after introducing NORAN to the transmitted signal. This is because NORAN increases the noise level in the EC, which reduces the capacity for the EC. On the other hand, the capacity for the LC remains unchanged. The inclusion of NORAN therefore increases the SC of the communication system.

In the case where the SNR of the eavesdropper remains constant, as the SNR of the LC decreases, the secrecy of the communication system shows a decreasing trend, as shown in Fig. 3(a). This is because a lower SNR implies a blurred distinction between signal and noise, leading to an increased error rate in demodulation at the receiver end. To counteract the noise interference, the communication system must reduce the data rate to maintain reliability. Therefore, a lower signal-to-noise ratio results in a reduction in the confidentiality of the communication system.

In the case where the SNR of the LC remains constant, the SC of the communication system decreases as the SNR of the EC increases, as shown in Fig. 3(b). This is because a higher SNR in the EC allows the eavesdropper to receive and decode the transmitted information more effectively, thereby increasing his capabilities and opportunities. Therefore, to ensure the security of the communication, it is important to minimise the SNR of the EC in order to limit the capabilities of the eavesdropper. In addition, maximising the inclusion of NORAN in the EC can effectively improve the security performance of the communication system.

Another key metric is the BER in communications, which represents the probability of bit errors occurring during data transmission.

After the introduction of NORAN, the BER of the eavesdropper experiences a certain increase. In addition, as shown in Fig. 2(b), we observed that the eavesdropper's BER shows an exponential growth trend with the change in SNR. This means that even a small decrease in SNR can result in a significant increase in BER. Therefore, maintaining a higher BER for the eavesdropper requires the eavesdropper to have a lower SNR. This is essential to ensure the security of the communication.

The analysis of Fig. 2(a), Fig. 3(a), and Fig. 3(a) reveals that there is no direct correlation between the number of transmitting and receiving antennas and the secure capacity of a communication system. In Fig. 2(a), the highest value of secure capacity is achieved when there are only two antennas, whereas in Fig. 3(a) and Fig. 3(b), the maximum secure capacity is observed when the antenna count is eight. This observation can be attributed to the inherent stochasticity of the channel, wherein increasing the number of antennas enhances the available channels and the independence among the transmitted signals, thereby improving the overall system capacity from a statistical perspective.

However, in practical scenarios, it is important to note that the presence of a greater number of antennas in a MIMO channel does not guarantee the optimal gain coefficient for signal transmission in a specific communication instance. Consequently, it cannot be inferred that the MIMO system with the highest antenna count during a particular communication instance will consistently exhibit the maximum secure capacity. Furthermore, the optimization algorithm employed for codebook generation, which is typically non-convex in nature, introduces an additional layer of complexity. The choice of initial values in algorithms like NORAN significantly affects the optimization performance, thereby introducing potential discrepancies between the secure capacity and the number of antennas in the MIMO system.

\section{conclusion}
In conclusion, this letter has addressed a novel anti-artillery noise - NORAN - designed on the basis of pilot information codebooks. The receiver of the information uses the pilot information as an index to query the codebook in order to mitigate the effects of NORAN. First, we provide a theoretical analysis highlighting the differences between NORAN based on pilot information codebooks and traditional NORAN. We show that effective design schemes for NORAN can be applied regardless of the prevailing channel conditions. Finally, through numerical simulations, we demonstrate that the incorporation of NORAN in wireless communication systems results in a certain improvement of the secure capacity without compromising the BER.

%
%

%
%



%

\end{document}